\def\nu{{\em Nuclear Fusion}}

\documentclass{ws-mpla}
\usepackage{amsfonts}
\usepackage{amssymb,amsmath}
\usepackage[super]{cite}
\usepackage{graphicx}
\begin{document}


\title{Electron Two-stream Instability and Its Application in Solar and Heliophysics }

\author{\footnotesize Haihong Che}

\address{ Department of Astronomy, University of Maryland, College Park, MD, 20742, USA\\
hche@umd.edu\\
and\\
Heliospheric Physics Laboratory, Goddard Space Flight Center, NASA, Greenbelt, MD, 20771, USA
}

\maketitle

\pub{Received (Day Month Year)}{Revised (Day Month Year)}

\begin{abstract}
It is well known that electron beams accelerated in solar flares can drive two-stream instability and produce radio bursts in the solar corona as well as in the interplanetary medium. Recent observations show that the solar wind likely originates from nanoflare-like events near the surface of the Sun where locally heated plasma escapes along open field lines into space. Recent numerical simulations and theoretical studies show that electron two-stream instability (ETSI) driven by nanoflare-accelerated electron beams can produce the observed nanoflare-type radio bursts, the non-Maxwellian electron velocity distribution function of the solar wind, and the kinetic scale turbulence in solar wind. This brief review focus on the basic theoretical framework and recent progress in the nonlinear evolution of ETSI, including the formation of electron holes, Langmuir wave generation in warm plasma, and the nonlinear modulation instability and Langmuir collapse. Potential applications in heliophysics and astrophysics are discussed.
\keywords{Beam Instability; Nonlinear Effects; Heating and Emission.}
\end{abstract}

\ccode{PACS Nos.:52.65.Rr,52.35.Qz,52.35.Mw}

\section{Introduction}
Electron Two-stream instability (ETSI) driven by electron beams was first discovered by Bohm and Gross in 1949\cite{bohm49pr}. The instability has since been found as one of the most common and important electrostatic instability in experimental beam plasma and space plasma. ETSI produces strong electron heating and coherent plasma emission at electron plasma frequency or harmonic. The emission frequency drifts as the plasma density varies along the path of electron beams. In the solar corona and solar wind, the emissions are in radio band, such as the  corona Type U and J radio bursts, and interplanetary Type III radio bursts \cite{wild54nat,wild63araa,lin73apjl,ergun98apj,aschw02ssr,aschw05book,beltran15sol}. Such transient coherent plasma emissions are also expected to be produced in the magnetosphere of neutron stars, giant planets, as well as in accretion disks around compact objects such as white dwarfs, neutron stars and blackholes\cite{ginzburg75araa,melrose03ppcf,coe14aa,kara15MNRAS}. The plasma emission carries information about the plasma around distant astronomical objects and the particle acceleration processes, and therefore is a powerful diagnostic tool in heliophysics and astrophysics. 

The study of ETSI drew tremendous interest among the heliophysics community after Wild discovered the solar radio bursts in 1954\cite{wild54nat,wild63araa},  and Ginzburg first related the radio emission to ETSI through nonlinear three-wave coupling  in 1959 \cite{ginzburg59book}. Interests in ETSI lasted for about twenty years and gradually waned.

The revival of the interest in ETSI came only recently, as increasing observational evidence supports the existence of coronal nanoflares, and as ETSI may play an important role in nanoflare heating and the formation of solar wind. High spatial and spectral resolution observations of the Sun from SOHO/SUMER and TRACE together with Extreme Ultraviolet Imaging Telescope (EIT) have revolutionized our view of the origin of solar wind. Different from the steady fluid solar wind model\cite{parker58apj,parker65ssr,jock70aap}, new observations\cite{axford77book, deforest97sol,zur99ssr, wang03apj,gloeckler03jgr,fisk03jgr,woo04apj,feldman05jgr,marsch07esa} found that the solar wind closely associates with impulsive events close to the surface of the Sun, which suggests that the solar wind originates from small magnetic loops rooted in the photosphere and escapes along open field lines caused by the merging of loops through granular convection\cite{wangym90apj,fisk01apj,xia03aap, feldman05jgr,tu05sci,brosius14apj}. 

The emerging dynamic picture of solar wind links nanoflares and solar wind. By nanoflare we refer to small scale explosive events that occur everywhere in the quiet sun, including corona holes. Our use of the term is an extension of Parker's original definition\cite{parker88apj}. The estimated occurrence rate of such events is $\sim 10^6$~$\rm{s}^{-1}$ for the whole Sun. Recent high resolution observation of the Sun from sounding rockets, spacecrafts SDO, IRIS and NuSTAR are providing increasingly detailed pictures of nanoflares\cite{win13apj,viall13apj,huang14apj,testa14sci,hannah16apj,klim15RSPTA}. Similar to flares, nanoflares can accelerate particles and the characteristic energy of nanoflare-accelerated electrons is in keV range\cite{gon13apj}. The accelerated electron beams can trigger ETSI, generate Langmuir waves, and produce type III radio bursts. Indeed, observations\cite{theja90sol,saint13apj} have found in the solar corona a new kind of type III radio burst whose brightness temperatures are about 9 orders of magnitude lower than flare-associated Type III bursts and are far more abundant, implying the bursts very possibly originate from nanoflares. The high occurrence rate of these ``nano type III bursts" indicates that electron beams and ETSI are common in the solar corona. The energetic charged particle streams produced by nanoflares are very likely source of the free energy that generates both the solar wind kinetic turbulence\cite{che14prl} and non-Maxwellian electron velocity distribution function (VDF)\cite{che14apjl}. 

How does the ETSI driven by nanoflares shape the kinetic properties of the solar wind? 
With the fast progress in high performance computers, particle-In-Cell simulations (PIC) which trace the motion of particles and fields and their interactions, is becoming a powerful tool to obtain deep understanding of nonlinear problems in plasma physics. Recent massive PIC simulations found that the nonlinear effects of ETSI can leave detectable kinetic signatures in the solar wind, such as the properties of electron VDF and the associated kinetic scale turbulence, and can naturally explain the origin of the observed non-thermal equilibrium in the solar wind and the kinetic scale turbulence and possibly nanoflare heating\cite{che14prl,che14apjl}. Coherent plasma emission produced by ETSI is found to be able to last for more than 5 orders of magnitude longer than its linear saturation time and this long duration possibly resolves the so called ``Sturrock dilemma"\cite{sturrock64}.  

This review focus on the key physical concepts and recent progress in the nonlinear evolution of ETSI. First in \S~\ref{overview}, a brief overview of the well established theoretical framework of ETSI is described. In \S~\ref{linear}, the kinetic theory of the ETSI and Langmuir wave generation by ETSI are summarized. In \S~\ref{mod}, modulation instability and Langmuir collapse are discussed. These are essential processes in nonlinear three-wave coupling. In \S~\ref{myresults}, recent progress in ETSI and its application in unifying the origin of kinetic properties of solar wind are presented. Finally, in \S~\ref{dis}, broad applications of ETSI in heliophysics and astrophysics are discussed.
\section{Overview of Electron Two-Stream Instability }	
\label{overview}
ETSI is an electrostatic instability whose wave vector $\mathbf{k}$ is parallel to the electric field fluctuation $\delta \mathbf{E}$, i.e. $\mathbf{k}\times \delta \mathbf{E}=0$. ETSI can be triggered when the relative drift between electron beams is larger than the electron thermal velocity. ETSI was studied under different physical conditions, such as different temperature, drift and density ratio of beams. ETSI demonstrates different evolution paths with different conditions. In cold plasma where phase speed of wave $v_p$ is much larger than the background electron thermal velocity $v_{te}$, ETSI can be well described by cold electron fluid equation in which pressure is neglected. The linear theory gives the frequency and growth rate of eigenmodes of electric field fluctuations. Once the magnitude of electric field fluctuations is large enough and the electron trapping becomes important, the linear theory becomes invalid and coherent structures, i.e. electron holes or solitons form\cite{bernstein57pr}. The trapping of electrons is a kinetic process and kinetic treatment is needed. In warm plasma where $v_p \sim v_{te}$, the effects of temperature must be considered. In this case the wave-particle interaction is strong and full kinetic description is needed to account for Landau damping. In warm plasma, nonlinear wave-wave and wave-particle couplings dominate the dynamics. Modulation instability and Langmuir collapse are two essential non-linear physical processes.

In heliophysical and astrophysical acceleration processes, it is common that the drift of the electron beams is much larger than the thermal speed, and the phase speed of waves generated by ETSI is much larger than the thermal speed too since the phase speed is proportional to the drift. Thus the cold plasma condition is commonly satisfied in space plasma. In the following we start from cold plasma to discuss different phases of ETSI as shown in Fig.\ref{etsichart}. In realistic conditions, ETSI can be triggered in either cold plasma or warm plasma and ETSI can end at any one of these stages depending on the initial temperature and kinetic energy of the beams.
\subsection{Linear growth phase---nonlinear growth phase---saturation of ETSI: the transition from cold to warm plasma }
Under cold plasma condition $v_p>>v_{te}$, the resonance occurs at the tail of the electron VDF and Landau damping is weak. The electron VDF can be approximated as a $\delta$ function (see \S~\ref{linear}).  In the cold plasma limit the phase speed of the fastest growing mode of ETSI $v_p=(n_b/2n_0)^{1/3}v_d$, where $v_d$ is the drift of electron beams, $n_0$ is the background electron density and $n_b$ is the beam density. Thus under cold plasma condition $v_d>(2n_0/n_b)^{1/3}v_{te}>v_{te}$, i.e. $w_{beam}>(2n_0/n_b)^{2/3} w_{te}$, where $w_{beam}= m_ev_d^2/2$ and $w_{te}=m_ev_{te}^2/2$---this condition can be satisfied in most space and astrophysical events. For example, in the solar flares, the plasma temperature is about 100~eV, and the typical energy for the solar flare accelerated electrons $w_{beam}\sim$~MeV, and the electron beam density is comparable to the background's\cite{dennis11ssr}. 
\begin{figure}
\includegraphics[scale=0.4,trim=10 270 30 10,clip]{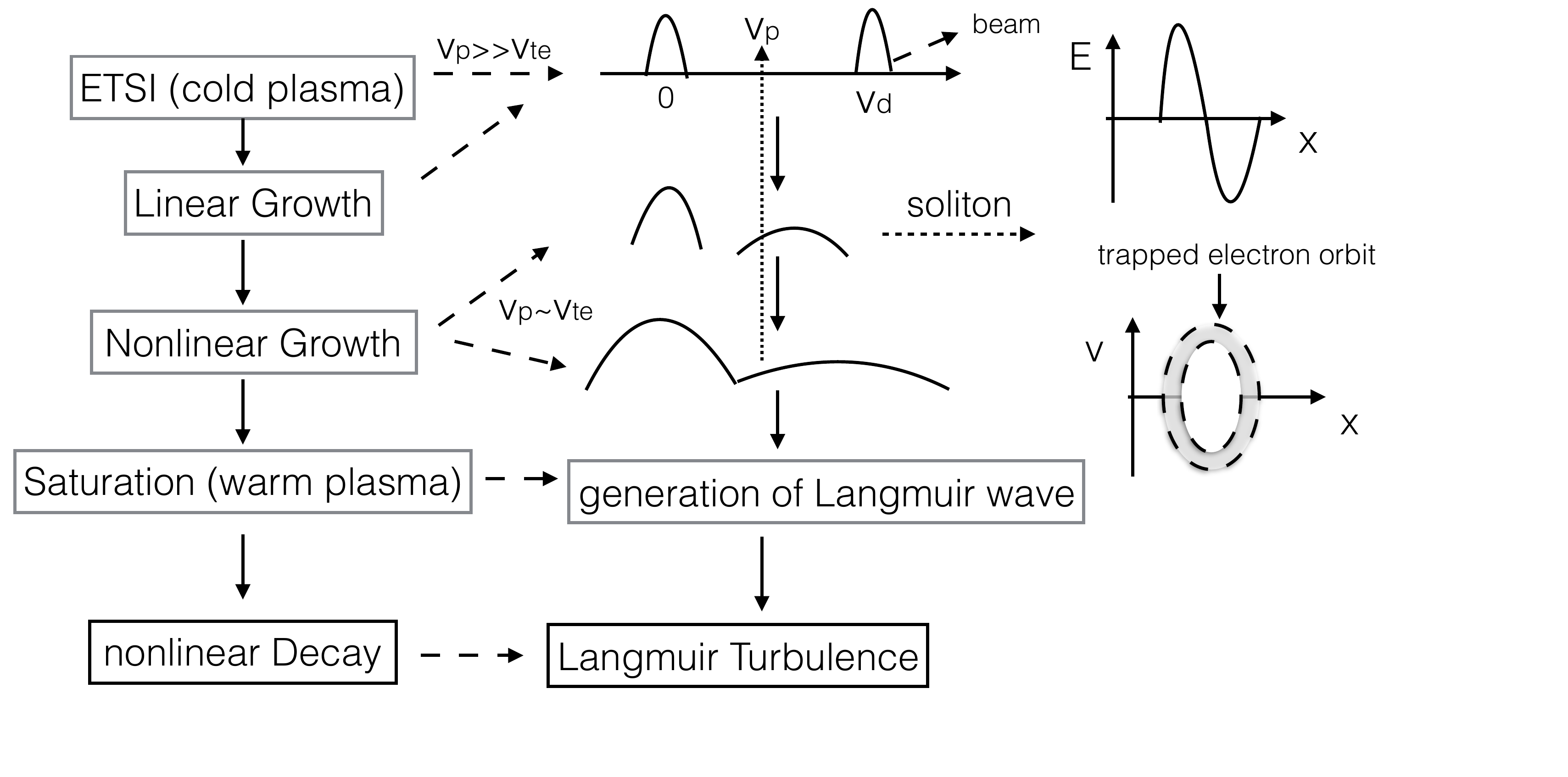} 
\caption{The evolution phases of ETSI. From the linear growth to the nonlinear decay. The electron heating leads to wider electron VDFs and warmer plasma. The kinetic effects dominate at nonlinear phases. }
\label{etsichart}
\end{figure} 

The linear growth rate $\gamma$ is of the same order as the electron plasma frequency $\gamma\sim \sqrt{3}/2(n_b/2n_0)^{1/3}\omega_{pe}$ and the fastest growing mode is $k_f= \omega_{pe}/v_d$. During the linear growth, most of the kinetic energy of the beams is converted into the growth of electric field $\delta E\propto e^{\gamma t}\sim ek_f m_e (v_{d}-v_{te})^2/2$. The linear growth time scale $1/\gamma$ is short and comparable to $1/\omega_{pe}$. 

If $v_d>2v_{te}$ and $\delta E>ek_f m_e v_{te}^2/2$ then the electric field can trap more electrons with velocity $<v_{te}$ and continue to grow. Consequently electron holes can develop. Kinetic theory is needed for the non-linear growth of ETSI and the formation of electron hole\cite{bernstein57pr,drum70pof}. The size of the coherent structure is close to the wavelength of the fastest growing mode of ETSI $\lambda\sim 1/k_f= v_d/v_{te} \lambda_{De}$, where $\lambda_{De}=v_{te}/\omega_{pe}$ is the electron Debye Length. The trapped electrons oscillate inside holes are governed by the energy conservation $m_e v^2/2+\phi(\mathbf{x})=const$. The electron hole is stable if its lifetime is much longer than electron bounce time $\sim \omega_b^{-1}$ where the bounce frequency is $\omega_b \sim k_f v_{b}$, and the electron bounce velocity $v_b$ is approximated as $v_b=(e\delta E_{max}/m_e k_f)^{1/2}$. The electron VDF inside a stable electron hole approximately satisfies $e^{-e\phi/2m_e v_{te}^2}$ and a phase space hole forms (see Fig.~\ref{etsichart}). 

The adiabatic motion of trapped electrons quickly exchange energy between the electrons and waves\cite{che13pop} and the coarse grain leads to phase mixing\cite{sagdeev69book}. This leads the electric fields to decay and the trapped electrons with higher energy break the confinement of electron holes and escape. The non-adiabatic de-trapping of electrons is irreversible\cite{che13pop}. Eventually these coherent structures are destructed and transform the cold beam into a long hot tail and the plasma becomes warm. The increase of the electron temperature is approximately balanced by the loss of the kinetic energy of electron beams $k\bigtriangleup T_e=n_b m_e \bigtriangleup v_d^2$. Meanwhile, the ETSI slowly transits from saturation to nonlinear decay phase (see Fig.~\ref{etsichart}).

\subsection{Saturation--Langmuir Turbulence--Plasma emission}
With the breaking of cold plasma condition, plasma is warm and the pressure becomes important. The density fluctuations of ions and electrons can generate waves on different scales. For example, ion density fluctuations produce low frequency (lower than ion plasma frequency) and long wavelength ion acoustic wave (IAW) while electron density fluctuations produce high frequency short wavelength Langmuir waves (at about the electron plasma frequency)\cite{oneil68pof,roberson71pof,che16book}. In warm plasma, Landau damping is sensitive to the temperature, density and drift. In nearly isothermal plasma, IAW is heavily damped and whistler waves takes over (see \S~\ref{myresults}). The generation of Langmuir wave is essential to the production of plasma coherent emission. Details on the generation of Langmuir wave in ETSI is presented in \S\ref{linear}.

The plasma emission is generated through the coupling between high frequency Langmuir wave and low frequency wave, such as ion IAW in plasma with high electron-ion temperature ratio\cite{zak72sjetp}. Recent PIC simulations show  that whistler wave dominates the coupling with Langmuir wave during the nonlinear decay of ETSI in isothermal solar corona plasma\cite{che16book}. The spatial disparate wave coupling drives modulation instability and transfers the energy from small scale to large scale, and develop a self-focusing Langmuir wave envelope where the plasma density depletes, also known as  ``\textit{caviton}"\cite{zak72sjetp,rud78physrep}. The wave spectra of inverse energy cascade is far away from the Landau damping region. Eventually the nonlinear process ``\textit{Langmuir collapse}" overrides this process as the magnitude of Langmuir wave envelope exceeds the critical threshold\cite{zak72sjetp}. 

\section{ Kinetic Theory of Electron Two-stream Instability in Warm Plasma and the Generation of Langmuir wave}
\label{linear}
ETSI shows different physical evolutions in cold plasma and warm plasma. To illustrate the difference and connections between these two cases, we focus on the Landau damping effect in ETSI using quasi-linear approximation of Vlasov equation in both the cold and warm homogeneous plasma\cite{bohm49pr,mik75book,anderson01ajp}. We will show that the Langmuir wave is generated in warm plasma when the drift of electron beam is comparable to its temperature \cite{oneil68pof,roberson71pof, mik75book}.

Let K be the reference frame moving with the denser plasma beam, i.e. the background plasma. In the frame K$^{\prime}$ moving at velocity $\mathbf{V}$ relative to K, the frequency $\omega$ is Doppler-shifted by $\omega^{\prime}=\omega-\mathbf{k}\cdot \mathbf{V}$, where $\mathbf{k}$ is the wave vector. Let's also assume that ions are motionless on electron time scale and neglect the ion influence on ETSI, the initial electron VDF including two beams can be written as $f_0=f_{1}+f_{2}$, the quasi-linear theory then gives the following dispersion relation: 
\begin{equation}
1=-\frac{4\pi e^2}{m_e k^2}\int^{\infty}_{-\infty}\frac{\partial f_0/\partial v}{\omega/k - v}dv.
\label{pequ}
\end{equation}

The cold plasma limit of ETSI can be obtained by assuming both $f_1$ and $f_2$ as $\delta$-functions $f_1=n_1 \delta (v)$ and $f_2=n_2 \delta (v-v_d)$, where  $n_1$is the density of denser electron beam 1 (background), $n_2$ is the density of electron beam 2, and $v_d$ is the relative drift of the electron beams. The dispersion relation then becomes:
\begin{equation}
1=\frac{\omega_{pe,1}^2}{\omega^2}+\frac{\omega_{pe,2}^2}{(\omega-kv_d)^2},
\label{gdis}
\end{equation}
where the electron plasma frequencies of two beams satisfy $\omega_{pe,1}/\omega_{pe,2}=n_1/n_2$. The growth rate for the fastest growing mode is $\gamma=\sqrt{3}/2(n_2/2n_1)^{1/3}\omega_{pe,1}$ and the corresponding wavenumber is $kv_d=\omega_{pe,1}$ and the real part of the frequency is $\omega_r=(n_2/2n_1)^{1/3}\omega_{pe,1}$. The corresponding phase speed is $v_p=\omega_r/k=(n_2/2n_1)^{1/3}v_d$\cite{chephd}. 

The cold fluid approximation of ETSI neglects Landau damping. To include the kinetic effects, we must consider the influence of the temperature of electrons. Assume the two electron beams have temperature $T_1$ and $T_2$, and the corresponding thermal velocity $v_{t1}\equiv \sqrt{T_1/m_e}$ and $v_{t2}\equiv\sqrt{T_2/m_e}$, the one dimensional VDFs can be written as 
 \begin{gather*}
 f_1=\frac{n_1}{\sqrt{2\pi}v_{t1}}  e^{-v^2/2v_{t1}^2},
f_2=\frac{n_2}{\sqrt{2\pi}v_{t2}}  e^{-(v-v_d)^2/2v_{t2}^2}.
\end{gather*}
The integral in the dispersion relation in Eq. (\ref{pequ}) is singular at $v=\omega/k$. This integral includes two contributions. One is from the principle value (PV) due to the main body of the distribution and the other is from the residue (Res) due to the electron resonance with wave at $v=\omega/k$. This gives
\begin{equation}
1=\frac{4\pi e^2}{m_e k}(PV\int_{-\infty}^{+\infty} \frac{\partial f_0/\partial v}{kv-\omega} dv + i\pi Res \int_{-\infty}^{+\infty} \frac{\partial f_0/\partial v}{kv-\omega} dv).
\label{reson}
\end{equation}
Expanding the denominator at $v=0$ when evaluating PV yields 
\begin{equation}
PV=-\frac{1}{\omega}\int_{-\infty}^{+\infty} (\frac{1}{\omega}+\frac{kv}{\omega^2} +\frac{k^2v^2}{\omega^3}+...)\frac{\partial f_0}{\partial v} dv.\nonumber
\end{equation}
Then we approximate the integral to the seconder order and  $PV=\omega_{pe,0}^2/\omega^2+3\omega_{pe,0}^2k^2v_{t0}^2/\omega^2$, where $\omega_{pe,0}^2=4\pi (n_1+n_2) e^2/m_e$, and $v_{t0}=\sqrt{n_1/n_0} v_{t1}+\sqrt{n_2/n_0} v_{t2}$ is the mean thermal velocity. The residual integral is $Res\int_{-\infty}^{+\infty} \frac{\partial f_0/\partial v}{v-\omega/k} dv=\frac{\partial f_0}{\partial v}\vert_{v=\omega/k}$.  Then the dispersion relation can be written as 
\begin{equation}
1=\frac{\omega_{pe,0}^2}{\omega^2}(1+\frac{3k^2v_{t0}^2}{\omega^2}) +
i\pi \frac{4\pi e^2}{m_e k^2}(\frac{\partial f_1}{\partial v}\vert_{v=\omega/k}+\frac{\partial f_2}{\partial v}\vert_{v=\omega/k})
\end{equation}

If the thermal and the resonant terms are small to the lowest order, the dispersion relation is reduced to $\omega\approx \omega_{pe,0}$. Using this approximation perturbatively we simplify the dispersion relation as:
\begin{equation}
\frac{\omega^2}{\omega_{pe,0}^2}=1+3k^2\lambda_{D0}^2 +
i\pi \frac{4\pi e^2}{m_e k^2} (\frac{\partial f_1}{\partial v}\vert_{v=\omega/k}+\frac{\partial f_2}{\partial v}\vert_{v=\omega/k}),
\end{equation}
where $\lambda_{D0}=v_{t0}/\omega_{pe,0}$.

Splitting frequency $\omega=\omega_r +i\gamma$ where $\gamma<<\omega_r\sim \omega_{pe,0}$, we obtain the dispersion relation for Langmuir wave, the high frequency electrostatic electron plasma wave in a warm plasma including Landau damping:
\begin{gather}
\label{real} \frac{\omega_r^2}{\omega_{pe,0}^2}=1+3k^2\lambda_{D0}^2,\\
\label{im} \frac{\gamma}{\omega_{pe,0}}
 =\sqrt{2\pi} (-\frac{\omega_{pe,1}^2 \omega_r}{k^3 v_{t1}^3}e^{-\omega_r^2 / 2k^2 v_{t1}^2}+\frac{\omega_{pe,2}^2 (v_d-\omega_r /k)}{k^2 v_{t2}^3}e^{-(\omega_r/k-v_d)^2/2v_{t2}^2}).
 \end{gather} 

Eq.~(\ref{real}) is the classical dispersion relation of Langmuir wave in a warm plasma. 
The negative sign in the first term of Eq.~(\ref{im}) means that the interaction between the background plasma and wave can only produce Landau damping. The second term on the other hand causes inverse Landau-damping due to $\partial f_2/\partial v >0$ if $v_d>\omega_r/k\sim v_{t0}\approx v_{t1}$, i.e. the threshold of ETSI.  The Langmuir wave will grow if $v_d\sim v_{t2}>>v_{t1}$ since the inverse Landau-damping suppresses the Landau damping. The relevant growth rate $\gamma$ is proportional to $1/k^2$, in other words long wavelength Langmuir wave grows faster --- this is the well studied gentle bump-in-tail instability\cite{drum64aop,oneil68pof,roberson71pof}. If $v_d>>v_{t2}$, the imaginary part can be neglected and dispersion relation reduces to the cold plasma limit in which the classic Langmuir wave will be replaced by the cold plasma ETSI.

\section{Modulation Instability and Langmuir Collapse in Warm Plasma}
\label{mod}
Modulation instability and Langmuir collapse are two essential processes in the nonlinear theory of wave-wave and wave-particle interactions, similar to the role Landau damping plays in the linear theory\cite{kad65book,sagdeev69book,sagdeev79rmp,krommes02pr,diamond14book}. These two processes have been extensively studied in the electron beam experiments\cite{cheung82prl,wong84prl,janssen84pof,jan84pof,cheung85prl},
simulations\cite{deg75sjetpa,deg75sjetpb,rowland77prl,pere77pof,zak89sjetp,newman90pop} and solar wind observations\cite{kellogg92grl,bale96grl,ergun08prl,theja13jgr}, and the results agree with theoretical predictions. The concepts and governing equations of modulation instability and Langmuir collapse are laid out in this section.  

Different from neutral fluid, plasma turbulence is characterized by several distinct scale lengths. For example, the ion gyroradius and electron gyroradius define the intrinsic length in magnetized plasma. The electron inertial length $c/\omega_{pe}$ is the scale of magnetic perturbation screening while the Debye length $\lambda_{De}=v_{te}/\omega_{pe}$ gives the boundary for collective collision. These scales are disparate due to the evident reason that the electron mass and ion mass differ substantially.

The presence of multi-scale waves in plasma allows a new class of multi-scale nonlinear interaction, i.e. disparate scale wave coupling. The wave with high frequency and small scale provides a ``source" for the wave with low frequency and large scale, through inducing a stress which acts upon them, as fluxes in their density, momentum and energy. The large scale wave acts as a ``strain" field in the presence of which the small scale wave evolves. Such interactions lead to the formation of large scale structure (flows, density and modulation, etc), a process similar to the ``inverse cascade" in fluid dynamics. The structures break once the amplitude of the structures reach the threshold, a process similar to ``forward cascade". However, in contrast to fluid turbulence energy cascade, the energy transfer does not occur through a sequence of intermediate scales, but rather proceeds directly between small and large scales. That's the reason why disparate scale wave coupling is an efficient mechanism for energy dissipation.

\begin{figure}
\includegraphics[scale=0.5,trim=50 230 60 80,clip]{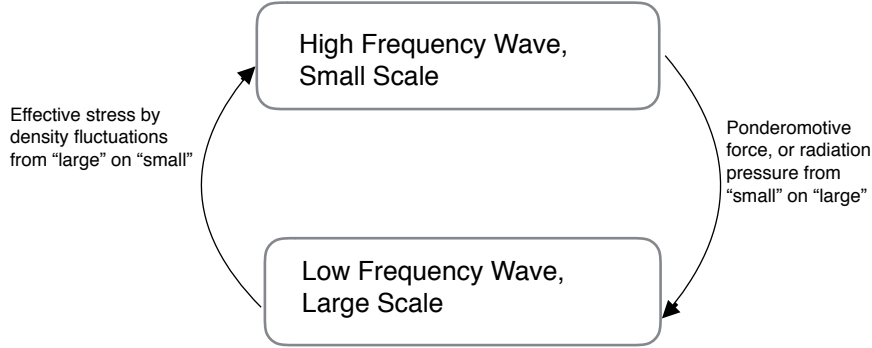}
\caption{Interaction of disparate scale wave coupling. }
\label{chart}
\end{figure}

The simplest and well-understood such interaction is the interaction between Langmuir waves and IAWs\cite{zak72sjetp,robinson97rmp}. In this case, 
high frequency langmuir waves exerts a low frequency ponderomotive force on electrons. The result ponderomotive pressure modulates Langmuir waves at large scale and induces the depletion of plasma density and the formation of caviton associated with long Langmuir envelope (Fig.~\ref{chart}). On the other hand, the density fluctuations associated with IAW causes refraction of Langmuir waves, so that the plasma waves tend to accumulate in an area of lower density. The amplitude modulation of Langmuir envelope grows and can lead to the sudden collapse of the modulated wave, i.e. the Langmuir collapse when the critical condition in Eq. (\ref{cond}) is satisfied.

Zakharov\cite{zak72sjetp} first mathematically described this process with two equations. The first equation describes how the density fluctuations $\delta n_i$ generated by IAW affect Langmuir wave. It is an equation on short scales:
\begin{equation}
\frac{i}{\omega_{pe}}\frac{\partial}{\partial t} E_L +\lambda_{De}^2\nabla^2 E_L^2=\delta n_i E_L,
\end{equation} 
where $E_L$ is the amplitude of Langmuir envelope. The second equation describes how the ponderomotive force $\nabla \langle E_L^2 \rangle /8\pi $ exerted by the Langmuir waves on electrons affects the Langmuir wave on large scale and interact with ions:
\begin{equation}
(\frac{\partial^2}{\partial t^2}-c_s^2 \nabla^2)\delta n_i=\frac{\nabla^2\vert E_L\vert^2}{4\pi n_0 m_i},
\end{equation}
where $c_s$ is the phase speed of IAW.
The growth rate for the modulation instability is $\omega_{pe} (m_e/m_i\vert E\vert^2/8\pi n_0 T_e)^{1/2}$ and critical condition for Langmuir collapse \cite{zak72sjetp,deg75sjetpa,deg75sjetpb} is
 \begin{equation}
\frac{\vert E_L\vert^2}{8\pi n_0 T_e}>\frac{1}{4}k^2\lambda_{De}^2.
\label{cond}
\end{equation}
Eq.~(\ref{cond}) tells us that the wavelength of Langmuir envelope determines the amplitude of Langmuir envelope, the longer the wavelength the weaker the electric field.  
 
More details on modulation instability and Langmuir collapse can be found in reviews by Rudakov \cite{rud78physrep} and Zakharov\cite{zak09pdnp}. A heuristic derivation was given by Sagdeev\cite{sagdeev79rmp} which gave a clear demonstration of the physical interaction between ion wave and Langmuir waves. A complete introduction to the subject can be found in the text by Diamond\cite{diamond14book}.
\section{The Detectable Relics of Nanoflare-Accelerated Electron Beams in the Solar Wind }
\label{myresults} 
\subsection{Two Long-standing Puzzles in the Solar Wind}
There are two long standing puzzles regarding the kinetic properties of the solar wind: 1) the origin and nature of kinetic scale turbulence in the solar wind, and 2) the origin of a nearly isotropic electron halo in the electron VDF. The two puzzles are thought to be closely related to the heating and acceleration of the solar wind\cite{ scudder92apja,scudder92apja,ko96grl,mak97aap,erdos12apj}. 
\begin{figure}
\includegraphics[scale=0.7,trim=50 510 130 10,clip]{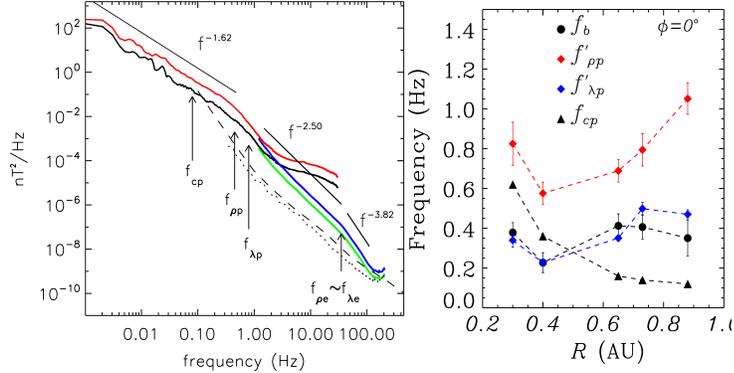}
\caption{{\bf Left:} Observation of solar wind magnetic turbulence power spectrum at 1AU\cite{sah09prl}. {\bf Right:} The evolution of magnetic power spectra break from 0.3-1AU\cite{bol12apjl}, where $f_b$ is the observed magnetic power spectral break frequency, $f_{\rho p,e}$ corresponds to the proton (electron) gyro-radius, and $f_{\lambda p,e}$ corresponds to the proton (electron) inertial length. The spectrum is observed in the solar wind moving frame.}
\label{break}
\end{figure}

Observations of solar wind turbulence (Fig.~\ref{break}) have shown that as scales approaching the ion inertial length where wave-particle interactions become important, the power-spectrum of magnetic fluctuations, which in the inertial range follows the Kolmogorov scaling law $B^2_{k} \propto k^{-5/3}$, is replaced by a steeper anisotropic scaling law $B^2_{k_\perp} \propto k_{\perp}^{-\alpha}$, where $\alpha > 5/3$. Spectral index  $\alpha\sim 2.7$ is found in observations but can vary between 2 and 4. Magnetic fluctuations with frequencies much smaller than ion gyro-frequency propagating nearly perpendicularly to the solar wind magnetic field are identified as  kinetic Alfv\'en waves (KAWs) \cite{Leamon_et_al_2000,bale05prl,sah09prl,kiyani09prl,salem12apjl,pod13solphys} and the break frequencies of the magnetic power-spectra from 0.3 to 1 AU suggest the break likely corresponds to the ion inertial length\cite{perri10apjl,bou12apj}. In the past decades, extensive studies of solar wind kinetic scale turbulence have focused on the idea that solar wind kinetic turbulence is due to cascade of large-scale turbulence. However, there are concerns that the energy in the solar wind large-scale turbulence may not be enough to cascade and support the observed kinetic scale turbulence and heating\cite{Leamon_et_al_1999}. 

\begin{figure}
\includegraphics[scale=0.9,trim=90 460 110 10,clip]{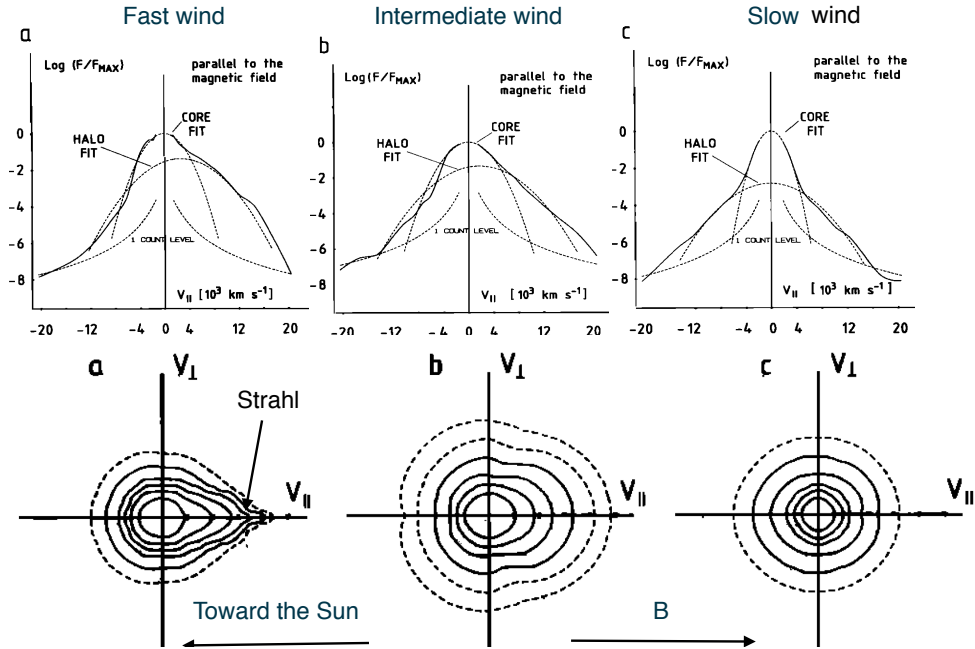}
\caption{The electron VDFs of solar wind at 1AU\cite{pilipp87jgra}. The top panels show the 1D cuts parallel to the magnetic field crossing the centers of 2D contours displayed below.  Slow wind is define with velocity $v_{wind} < 400$~km~s$^{-1}$ and fast wind with $v_{wind} > 700$~km~s$^{-1}$. The speed of the intermediate wind is  between these two. }
\label{pilipp}
\end{figure}

Observations of electron VDFs at heliocentric distances from 0.3 to 1 AU show a prominent ``break" or a sudden change of slope at a kinetic energy of a few tens of electron volts as shown in Fig.~\ref{pilipp}. The electron VDF below the break is dominated by a Maxwellian known as the ``core" while the flatter wing above the break is called the ``halo"\cite{pilipp87jgra,pilipp87jgrb}. So far no model can naturally produce the nearly isotropic halo population that can be described by a kappa function\cite{marsch06lrsp}. The isotropic nature of the halo suggests that halo formation needs strong turbulence scattering and is likely related to the kinetic turbulence in the solar wind\cite{marsch06lrsp}. In addition, ``strahl" -- an anisotropic tail-like feature skewed with respect to the magnetic field direction is found in the electron VDF of fast solar wind with speed $\geqslant 400$~km~s$^{-1}$.  In the slow solar wind coming from the sector boundary with speed $< 400$~km~s$^{-1}$, the strahl is nearly invisible and the isotropic core-halo feature dominates. Existing kinetic models based on magnetic focusing effect with input of different modes of kinetic turbulence, while successful in producing the strahl-like tail at a minimum heliocentric distance of 10 $R_{\odot}$\cite{smith_hm12apj,landi12apja}, are unable to produce the halo\cite{marsch06lrsp,marsch12ssr,vocks12ssr}. This suggests that extra energy dissipation, probably by plasma kinetic instabilities, is required \cite{marsch06lrsp,marsch12ssr}. 
At large heliocentric distances (0.3 - 1 AU), observations found the electrons are scattered by whistler waves from strahl to halo\cite{mak05jgr,stverak09jgr,gurg12ag}, and theories suggest that the whistler scattering might be caused by the unstable processes related to the strahl but unable to produce the isotropy of electron halo\cite{vocks05apj,mith12pop}. 

Clearly to generate the observed global isotropic halo and the observed turbulent fluctuations on kinetic scales, some form of  ``extra free energy", probably continuously launched from the Sun, is needed. The energy source should be uniformly distributed in the Sun and able to produce kinetic scale instability. The nanoflare accelerated electron beams in the corona is one energy source known to satisfy these conditions. These beams can release their kinetic energy through ETSI and produce the kinetic scale turbulence and plasma emission. We recently employed PIC simulations to investigate this possibility and have shown how ETSI could generate KAWs and whistler waves on kinetic scales\cite{che14prl,che14apjl}.

\subsection{PIC Simulations of ETSI in Nanoflares}
Recently 2.5D massive parallel PIC simulations are carried out to study the nonlinear evolution of ETSI in a uniformly magnetized plasma with equal ion and electron temperature\cite{che14prl,che14apjl,che16book}. The initial physical parameters resembles the typical physical condition in the solar corona. The initial drift of the electron beams is large compared to the thermal velocity so that the ETSI starts from a cold plasma and ends as warm plasma. 

\begin{figure}
\includegraphics[scale=0.6,trim=100 220 30 50,clip]{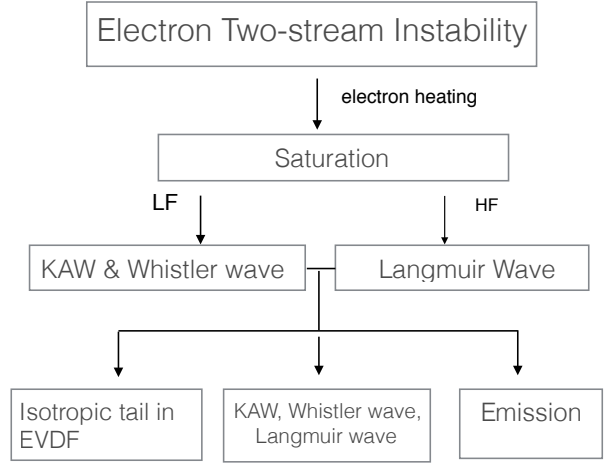} 
\caption{The main processes during the nonlinear evolution of ETSI. The following abbreviations are used: HF: high frequency, LF: low frequency, EVDF: electron VDF.}
\label{mainchart}
\end{figure}

As shown in Fig.\ref{etsichart}, the ETSI experience four phases: linear growth, nonlinear growth, saturation and nonlinear decay till turbulent equilibrium. The time scale of the linear growth phase is about 1/10000 of the total evolution period of ETSI while the linear and nonlinear growth time is 1\% of the total. The evolution of the ETSI spends most of the time in the nonlinear decay phase. The complete evolution of ETSI involves several new nonlinear processes as shown in (Fig.~\ref{mainchart}), including 1) the generation of low frequency KAWs and whistler waves through bidirectional energy cascades; 2) the generation of high frequency Langmuir waves and plasma emission through repeating Langmuir collapse coupling with low frequency kinetic waves. The wave-wave and wave-particle interactions eventually lead to the balance of energy exchange, and the turbulence stays at a non-thermal equilibrium in which the non-Maxwellain electron VDF and kinetic waves co-exist. These relics of the violent dissipation through ETSI can be carried out into interplanetary with the solar wind along open field lines\cite{che14apjl}. 
\subsection{ The Common Origin of Kinetic Scale Turbulence and Non-Maxwellian Electron VDF in the Solar Wind }
In traditional kolmogorov turbulence energy is injected from large scale, cascades into small scale and eventually is dissipated by viscosity. ETSI, however, injects beam kinetic energy on the scales close to Debye length, the energy cascade is quickly stopped by strong electron heating. On the other hand, we found that KAW and whistler wave coupling can inversely transfer the energy into large scale and develop kinetic scale turbulence. Simultaneously, the waves scatter the hot electron tail that lies along magnetic field into an isotropic population superposed over the Maxwellian core electron VDF, forming the electron halo in the solar wind\cite{che14prl,che14apjl}.
\begin{figure}
\includegraphics[scale=0.8,angle=0,trim=70 500 80 50,clip]{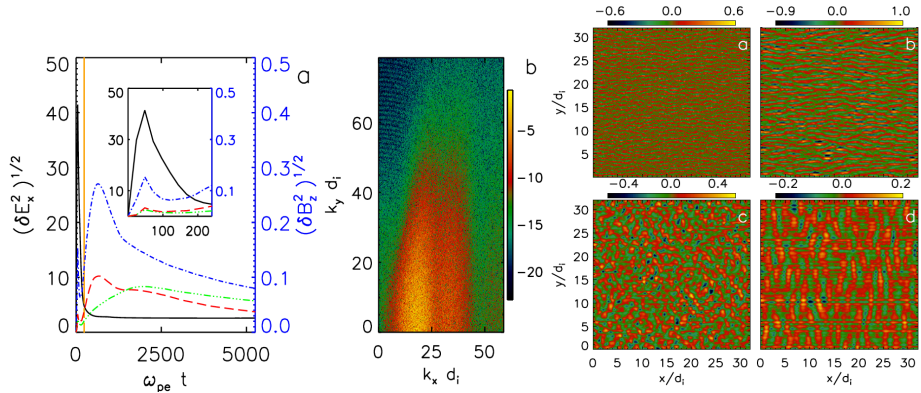} 
\caption{\textbf{Left panel:} Time evolution of turbulent energies: $\langle \delta E_x^2\rangle$ (black line), $\langle \delta B_x^2\rangle$ (red line), $\langle \delta B_y^2\rangle$ (green line), $\langle \delta B_z^2\rangle$ (blue line).  The embedded plot is an expanded view of the time evolution from $\omega_{pe} t=0-230$. The orange line indicates $\omega_{pe} t=230$. \textbf{Middle panel:} Power spectrum of $\vert \delta E_x(k_x, k_y)\vert^2$ at the linear phase of ETSI, shown in logarithmic scale. \textbf{Right panel:} $B_z/B_0$ at linear phase ({\bf a}), nonlinear growth phase ({\bf b}),  nonlinear decay phase ({\bf c}), and turbulence equilibrium ({\bf d}).}
\label{fluex}
\end{figure} 

The linear growth of ETSI lasts about $\omega_{pe}t=20$ (Fig.~\ref{fluex}), and around 10\% is converted into magnetic energy at the nonlinear growth phase, while nearly 90\% is converted into the thermal motion of trapped electrons. The fast growth of electric field induces a magnetic field, and the electric current density  $j_{ex}$ produced by the inductive magnetic field becomes as important as the displacement current when the ETSI starts to decay. The current $j_{ex}$ then drives a Weibel-like instability that generates nearly non-propagating transverse electromagnetic waves. The fast decay of the localized $j_{ex}$ breaks up the transverse waves and produces randomly propagating KAWs and whistler waves. The wave-wave interactions drive a bi-directional energy cascade. The perpendicular KAW energy is transferred from the electron inertial scale up to the ion inertial scale. The parallel whistler wave energy is transferred from the ion inertial scale down to the electron inertial scale. Eventually, magnetic power is concentrated in two branches in the energy spectrum: the nearly perpendicular branch with $k_x d_i<1$, and the parallel branch with $k_y d_i <2$ (see Fig.~\ref{fluex}). Around $\omega_{pe} t=10000$, the energy exchange between particles and waves reaches a balance. The turbulence reaches its new steady state with $P^2 + B^2/8\pi=constant$, where $P$ is the total pressure of ions and electrons.
\begin{figure}
\includegraphics[scale=0.8,angle=0,trim=70 550 80 50,clip]{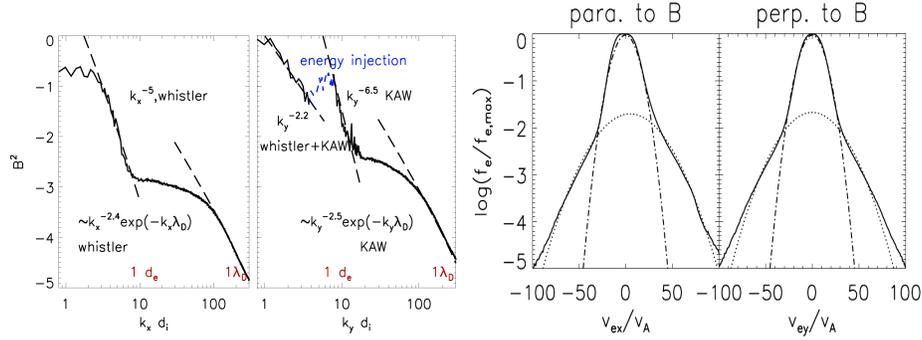} 
\caption{\textbf{Left panel:} The 1D spectra of $ \delta B^2(k)$ vs. $k_x d_i$  and $k_y d_i$. The blue short-dashed line shows the $k_y$ s range for magnetic energy injection. \textbf{Right panel:} The 1D electron VDF cuts parallel and perpendicular to magnetic field when ETSI simulation reaches turbulence equilibrium. The dot-dashed lines delineate the core Maxwellian VDF and the dashed lines represent the halo VDF. They are plotted in the same manner as in Pilipp et al. (1987).  }
\label{spec1d}
\end{figure} 

The amplitude ratio of the magnetic fluctuations to the background magnetic field is $\sim 0.2$, which agrees with observations of solar wind kinetic turbulence. The 1D magnetic fluctuation power spectrum is shown in Fig.~\ref{spec1d}. With $1<k_y d_i<2$, which corresponds to the range of wave lengths current instruments can probe, both KAWs and whistler waves are important. The perpendicular power spectrum is fitted with a power-law with an index of -2.2 which falls within the observed range. The perpendicular power spectrum terminates at the ion inertial length, is also consistent with observations\cite{Leamon_et_al_1999,perri10apjl,bou12apj}. A ubiquitous observable feature is the spectral break at the electron scale caused by energy injection. This model also predicts the existence of whistler waves, and the cutoff of the parallel power spectrum at the ion gyro-radius\cite{che14prl}. 

The steady state electron VDF in our simulation agrees with the observed core-halo structure in the solar wind (see Fig.~\ref{spec1d}). This is expected if the beam heated plasma escapes from the inner corona and advects into interplanetary medium along open field lines, forming the solar wind and preserving its kinetic properties. This nonlinear heating process predicts that the core-halo temperature ratio $T_h/T_c$ of the solar wind is insensitive to the initial conditions in the corona but is linearly correlated to the core-halo density ratio of the solar wind $n_c/n_h$:
\begin{equation}
\frac{T_h}{T_c}  \approx \frac{n_c}{n_h}\frac{1-C_T}{C_T}+4,
\label{dthc2}
\end{equation} 
where $C_T$ is the rate at which kinetic energy of electron beams converts to heat, and $C_T \sim 0.9$ is found in our simulations. If the core and halo experience similar temperature evolutions when travelling from the Sun to 1AU, the temperature ratio can be approximately preserved.  In fast wind where the strahl is strong,  the halo temperature can be replaced by the mean temperature of halo and strahl $T_{hot}=n_{strahl} T_{strahl}/n_c +n_h T_h/n_c$ and halo density be replaced by the total density of both strahl and halo $n_{hot}=n_{strahl}+n_h$ because the energy and density are approximately conserved during scattering\cite{mak05jgr}:
\begin{equation}
\frac{T_{hot}}{T_c}  \approx \frac{n_c}{n_{hot}}\frac{1-C_T}{C_T}+4,
\label{dthc3}
\end{equation} 
The break point dividing the core and halo in electron VDF, which is a useful quantity in observations, satisfies:
\begin{equation}
v_{brk}\approx[\ln(T_{h}/T_c)-\ln (n_h/n_c)^2]^{1/2} v_{te,c}.
\label{brk}
\end{equation} 
In addition, the relative drift between the core and halo is close to the core thermal velocity -- a relic of the ETSI saturation.

The simulations\cite{che14prl} show that when the kinetic turbulence fully saturates, the ratio of parallel to perpendicular electric field fluctuations $\langle \vert \delta E_{\parallel}\vert/\vert \delta E_{\perp}\vert\rangle$ is enhanced by the relic parallel electric field by a factor of $\sim 2-3$, consistent with observations that the parallel turbulent electric field is larger than the perpendicular turbulent electric field, contrary to what is expected if the turbulent fluctuations are dominated by KAWs \cite{mozer13apjl}. The enhanced electric field might be caused by electrostatic whistler wave and Langmuir wave. 

\subsection{Nonlinear Plasma Emission and the Sturrock Dilemma }
We have discussed that the linear growth phase of ETSI is very short but the nonlinear evolution phase of ETSI is significantly longer. The coupling between Langmuir wave (L) and whistler wave (W) leads to the growth of modulation instability. The energy on Debye scale is inversely transferred into ion scale. As a result, Langmuir wavepackets (cavitons) characterized by the density depletion are form. Plasma emission is produced through $L+W\rightarrow T$, where $T$ is transverse emission. As the modulation instability grows and the critical condition in Eq.(\ref{cond}) is satisfied, Langmuir collapse occurs. Our simulation shows that the plasma emission is continuously produced beyond the turbulence equilibrium. A new paper of ours that is to be submitted will concentrate on how nonlinear ETSI maintains a nonlinear feedback loop and produces long duration plasma emission.

In the simulations, the turbulence reached its non-thermal equilibrium in $\sim 10^4\omega_{pe}^{-1}$. Since the modulation instabilities nearly dominate the whole process, the nonlinear saturation time is approximately proportional to $(m_i/m_e)^{1/2}$ which translates to a turbulence saturation time five orders of magnitude longer than ETSI linear saturation time $(n_0/n_b)^{1/3} \omega_{pe}^{-1}\sim 2 \omega_{pe}^{-1}$ for real mass ratio. The simulation assumes the beam energy being $\sim 100$ times the coronal temperature. For nanoflares, the beam energy is in the keV range while the inner corona plasma temperature is 10 eV.  The beam can produce weak coronal radio Type III bursts with short duration\cite{mercier97apjl,saint13apj}. Our results can also be applied to flares. For ``normal" flares, the electron beam energy can reach MeV. While the ETSI growth rate does not depend on the velocity once the threshold is reached, the resulting turbulence however becomes stronger, the decay lasts longer and produce strong flare type III radio bursts. The mechanism we demonstrated provides a complete and self-consistent solution to the long-standing mystery of why the durations of solar radio bursts are much longer than the linear saturation time of ETSI, also known as the ``sturrock dilemma"\cite{sturrock64}.

To summarize, our recent work show that nanoflare-accelerated electron beams can cause ETSI which generates kinetic turbulence and the non-Maxiwellian electron VDF consistent with what is observed in the solar wind. Previous studies based on observations of the solar corona suggest that heated plasma in the inner corona can escape from open field lines and form the solar wind\cite{fisk03jgr,gloeckler03jgr}. If this is true this model provides a crucial heating mechanism. The major attraction of this new model is that it can account for the origin of both the electron VDF and kinetic turbulence in a unified manner, while past studies treat these two phenomena as unrelated. The link between the solar wind and nanoflares directly relates solar wind properties to photosphere dynamics, and puts useful constraints on kinetic processes in both the solar corona and solar wind. Further, nanoflares also accelerate ions, and ion dynamics is critically important to the formation and accelerations of solar wind. This requires us to incorporate ion beams into this model to understand the dynamics of ions as well as how ion and electron waves interact. This issue will be addressed with future studies. 
\section{ Potential Applications in Heliophysics and Astrophysics }
\label{dis}
The kinetic scale physics is becoming accessible experimentally thanks to the new generation of space probes and solar observatories. The newly launched {\it Magnetospheric MultiScale Mission} (MMS) is capable of measuring solar wind turbulence on electron scales at 1~AU, and the ongoing {\it Solar Probe Plus} (SPP) can conduct {\it in situ} measurements of the electron VDF of solar wind at a heliocentric distance as close as 10 R$_{\odot}$, and observe the nanoflares at such close distances. The plasma wave detectors onboard SPP and {\it Solar Orbiter} have the capability of observing nanoflare associated radio bursts. Electron beams can also affect Balmer line emission and produce X-ray and EUV emissions, and such effects can be observed by IRIS\cite{testa14sci} and NuSTAR\cite{hannah16apj}.

One important aspect of solar physics is to provide astronomers a better understanding of the sun-like stars\cite{brun15ssr}. It is essential for us to know of stellar convection, rotation and magnetism and to assess the degree to which the Sun and other stars share similar dynamical properties.  Various emissions and energetic particles we are able to detect are valuable. For example the observation of radio bursts produced by stellar flares is a powerful tool to probe the plasma properties and the magnetic fields of the stars.  

Besides normal star emissions, microsecond radio emissions from radio pulsars are believed to be plasma coherent emission, although there is a discrepancy between the plasma density expected from the radio emission and the standard model of the magnetosphere of pulsar\cite{goldreich69apj,melrose92rspta,melrose03ppcf,han03nat}. Different from the normal stars, pulsars are highly magnetized with magnetic field up to $10^{12}$~G while the magnetic field of solar corona is only 1~G. The pulsar plasma is highly relativistic electron-positron pair plasma in the source region\cite{sturrock71apj,melrose03ppcf,beskin15ssr}. The Langmuir collapse driven by relativistic beam instability is suggested to be responsible for the microsecond radio bursts\cite{melrose92rspta,wea97apj,han03nat}. 
\section*{Acknowledgments}
The author would like to thank her collaborators Drs. Melvyn Goldstein, Adolfo  Vi\~nas, Profs. Roald Sagdeev, and Patrick Diamond for the supports in this study. 

\bibliographystyle{ws-mpla}
\bibliography{solarwind}


\end{document}